\documentclass[twocolumn,showpacs,preprintnumbers]{revtex4}%
\usepackage{amssymb}
\usepackage{amsmath}
\usepackage{graphicx}
\usepackage{amsfonts}%
\providecommand{\U}[1]{\protect\rule{.1in}{.1in}}

\begin{document}
\title{Cosmological Particle Creation in the Presence of Lorentz Violation}
\author{E. Khajeh}
\thanks{e-khajeh@sbu.ac.ir}
\author{N. Khosravi}
\thanks{n-khosravi@sbu.ac.ir}
\author{H. Salehi}
\thanks{h-salehi@sbu.ac.ir}
\affiliation{Department of Physics, Shahid Beheshti University, Evin, Tehran 19839, Iran}

\begin{abstract}
In recent years, the effects of Lorentz symmetry breaking in
cosmology has attracted considerable amount of attention. In
cosmological context several topics can be affected by Lorentz
violation,e.g., inflationary scenario, CMB, dark energy problem
and barryogenesis. In this paper we consider the cosmological
particle creation due to Lorentz violation (LV). We consider an
exactly solvable model for finding the spectral properties of
particle creation in an expanding space-time exhibiting Lorentz
violation. In this model we calculate the spectrum and its
variations with respect to the rate and the amount of space-time
expansion.

\end{abstract}

\pacs{04.62.+v, 11.30.Cp, 98.80.-k, 98.80.Jk}
\maketitle
\thanks{The minus in $1^{-}$ means approaching $1$ from the left.}


\section{Introduction}

Particle creation in the cosmological context is one of the most interesting
feature of QFT in curved space time \cite{1}. In this process, a quantum field
propagates on an expanding space-time and the field quanta (particles) are
generated through an impulse of the space-time expansion. The density and the
rate of particle production depend on the vigor of the expansion.

Furthermore, it is well known that if a conformally invariant field propagates
in a space-time which is conformally equivalent to Minkowski (conformal
triviality) no particle production occurs. Thus, we expect that massless
quanta of Maxwell and Dirac fields do not arise from the expansion of
Universe. But if a term is added to the field equation such that conformal
invariance is broken then these particles are created \cite{1,1.5}.

QFT in curved background is based on the hypothesis that the field equations
are locally Lorentz invariant. But in recent years there have been many
attentions to the possibility of Lorentz symmetry breaking at high energies.
The initial motivation came from string theory \cite{2}, and more recently
this symmetry breaking has been discussed in the context of noncommutative
geometry \cite{3,3.5}. Also, there has been evidence that this symmetry may be
broken in at least three different phenomena:

i) Observation of ultra-high energy cosmic rays with energies beyond the
so-called GZK cutoff, $E_{GZK}\backsimeq10^{19}eV$ \cite{4,4.5}.

ii) Events involving gamma radiation with energies beyond 20TeV from distant
sources such as Markarian 421 and Markarian 501 blazers \ \cite{5}.

iii)Studies of the evolution of air showers produced by ultra high-energy
hadronic particles suggest that pions live longer than expected \cite{6}.
These observations can be explained via the breaking of Lorentz symmetry
\cite{6.5,6.75}. Now, if we want to consider the effect of LV in QFT in curved
space-time, an important question for investigation is: \textit{How dose
Lorentz violation affect the particle creation in an expanding space-time?}

The possible effects of LV physics in inflationary cosmology has been studied
in \cite{BM,EG,S,MB}. The authors found that the spectrum of fluctuations in
inflationary cosmology and so the spectrum of temperature anisotropies in the
Cosmic Microwave Background can be affected by LV. Also it is shown that
Lorentz violation is relevant to the dark energy problem and barryogenesis
\cite{14.5,14.75,14.85,14.95}. In the present work we look for the effects of
LV in particle creation process in an expanding space-time. Our interest is to
study the form and properties of particle creation spectrum of a scalar field.
For this purpose we choose an exactly solvable model and we find the
characteristics of the spectrum.

We benefit from the model that has been introduced in \cite{7,9} for LV to
study this subject. In this model the usual dispersion relation is modified by
adding a term $\alpha^{2}k^{4}$. The modified dispersion relation preserves
rotational invariance but violates the boost invariance of Lorentz symmetry.
It has been shown that adding a term $\alpha^{2}k^{4}$ to the dispersion
relation is equivalent to putting a vector field which is coupled with matter
field in the Lagrangian of the model. This vector field can physically be
assumed as the four velocity of a preferred inertial observer. The additional
term in the Lagrangian which enforces the LV, breaks the conformal invariance.
Therefore we expect that the massless field quanta are created during the
space-time expansion.

The organization is as follows: in section \ref{1010} we quantize the Lorentz
violation model introduced in \cite{7} on Minkowski space-time. In section
\ref{1011} we define an expanding cosmological model and consider the
quantization of the LV model on this space-time. Finally, we obtain and
discuss the spectrum characteristics of created massless particles in details.


\section{Lorentz Violation Model and Quantization in Minkowski space-time}

\label{1010}
As was mentioned before, in the model introduced for LV in \cite{7} the
dispersion relation is modified by an additional term $\alpha^{2}k^{4}$ and
the modified dispersion relation reads as follows%
\begin{equation}
\omega^{2}(\vec{k})=\left\vert \vec{k}\right\vert ^{2}-\alpha^{2}\left\vert
\vec{k}\right\vert ^{4}. \label{a}%
\end{equation}
A Lagrangian that can lead the above dispersion relation for a scalar field
$\phi$ is
\begin{equation}
\mathcal{L}=\frac{1}{2}(\partial^{\mu}\varphi\partial_{\mu}\varphi+\alpha
^{2}(D^{2}\varphi)^{2}), \label{b}%
\end{equation}
where $\alpha$ is a constant (of order the Planck energy) that sets the scale
of the Lorentz violation and $D^{2}$ is the spatial Laplacian, that is
\begin{equation}
D^{2}\varphi=-D^{\mu}D_{\mu}\varphi=-q^{\mu\nu}\partial_{\nu}(q_{\mu}^{\tau
}\partial_{\tau}\varphi), \label{b2}%
\end{equation}
where $q_{\mu\nu}$ is the (positive definite) spatial metric orthogonal to the
unit timelike vector $u^{\mu}$
\begin{equation}
q_{\mu\nu}=-\eta_{\mu\nu}+u_{\mu}u_{\nu}\hspace{0.5cm},\hspace{0.5cm}\eta
^{\mu\nu}u_{\mu}u_{\nu}=1. \label{d}%
\end{equation}
The vector field $u^{\mu}$ can physically be interpreted as the four velocity
of a preferred inertial observer. The rest frame of this preferred observer
may be called aether. In this rest frame we have $u^{\mu}$=$\left(
1,0,0,0\right)  .$

For the subsequent consideration in this chapter we shall take a coordinate
system in which $u^{\mu}$\ is constant (The aether is a particular example for
such a coordinate system). With this assumption the equation of motion for the
Lagrangian (\ref{b}) is
\begin{equation}
\lbrack\square-\alpha^{2}q^{\mu\nu}q^{\gamma\delta}\partial_{\mu}\partial
_{\nu}\partial_{\gamma}\partial_{\delta}]\varphi=0. \label{e}%
\end{equation}
The complex mode solution of (\ref{e}) are taken as
\begin{equation}
u_{\vec{k}}\propto e^{-ik_{\mu}x^{\mu}}, \label{f}%
\end{equation}
where
\begin{equation}
\omega^{2}(\vec{k})=|\vec{k}|^{2}-\alpha^{2}q^{\mu\nu}q^{\gamma\delta}k_{\mu
}k_{\nu}k_{\gamma}k_{\delta}. \label{g}%
\end{equation}

In general the relation (\ref{g}) admits imaginary frequencies. These
frequencies lead to instability and unboundedness \cite{1.5}. It is necessary
in this context to restrict ourselves to ordinary solutions with real
frequencies. In this case the mode solutions (\ref{f}) become positive
frequency mode solutions.

We define the scalar product for two solutions $\varphi_{1}$ and $\varphi_{2}$
of Eq.(\ref{e}) as follows
\begin{equation}
(\varphi_{1},\varphi_{2})=-i\int_{\Sigma}\varphi_{1}(x)(\overleftrightarrow
{\partial}^{\mu}-\alpha^{2}q^{\mu\nu}q^{\gamma\delta}\overleftrightarrow
{{\partial_{\nu}\partial_{\gamma}\partial}}{_{\delta}})\varphi_{2}^{\ast
}(x)d\Sigma_{\mu}, \label{h}%
\end{equation}
where $d\Sigma_{\mu}=n_{\mu}d\Sigma$, with $n_{\mu}$ a future-directed unit
vector orthogonal to the space like hypersurface $\Sigma$ and $d\Sigma$ is the
volume element in $\Sigma$. The hypersurface $\Sigma$ is taken to be a Cauchy
surface in the space-time and we can show, using Gauss' theorem, that the
value of $(\varphi_{1},\varphi_{2})$ is independent of $\Sigma$. The notation
$\overleftrightarrow{{\partial_{\nu}\partial_{\gamma}\partial}}{_{\delta}}$ in
(\ref{h}) is defined by%
\begin{equation}%
\begin{array}
[c]{l}%
\varphi_{1}\overleftrightarrow{{\partial_{\nu}\partial_{\gamma}\partial}%
}{_{\delta}}\varphi_{2}=\\
+\varphi_{1}{\partial_{\nu}\partial_{\gamma}\partial_{\delta}}\varphi
_{2}-\varphi_{2}{\partial_{\nu}\partial_{\gamma}\partial_{\delta}}\varphi
_{1}\\
-\partial_{\nu}\varphi_{1}\partial_{\gamma}\partial_{\delta}\varphi
_{2}+\partial_{\nu}\varphi_{2}\partial_{\gamma}\partial_{\delta}\varphi_{1}\\
+\partial_{\gamma}\varphi_{1}\partial_{\nu}\partial_{\delta}\varphi
_{2}-\partial_{\gamma}\varphi_{2}\partial_{\nu}\partial_{\delta}\varphi_{1}\\
-\partial_{\delta}\varphi_{1}\partial_{\gamma}\partial_{\nu}\varphi
_{2}+\partial_{\delta}\varphi_{2}\partial_{\gamma}\partial_{\nu}\varphi_{1}.
\end{array}
\label{n}%
\end{equation}
The ordinary modes $u_{\vec{k}}$ are orthogonal%
\begin{equation}
(u_{\vec{k}},u_{\vec{k}^{\prime}})=0\hspace{0.5cm},\hspace{0.5cm}\vec{k}%
\neq\vec{k}^{\prime}, \label{o1}%
\end{equation}
and if we choose
\begin{equation}
u_{\vec{k}}=[2(2\pi)^{3}(\omega-2\alpha^{2}q^{0\nu}q^{\gamma\delta}k_{\nu
}k_{\gamma}k_{\delta})]^{-\frac{1}{2}}e^{-ik_{\mu}x^{\mu}}, \label{j}%
\end{equation}
these ordinary modes are normalized in the sense of the scalar product
(\ref{h}). An ordinary solution of Eq.(\ref{e}) may be expanded in term of the
ordinary modes (\ref{j}) and their complex conjugates
\begin{equation}
\varphi(t,\vec{x})=\sum_{\vec{k}}a_{\vec{k}}u_{\vec{k}}(t,\vec{x})+a_{\vec{k}%
}^{\dagger}u_{\vec{k}}^{\ast}(t,\vec{x})\hspace{0.8cm}\omega\in%
\mathbb{R}
, \label{k}%
\end{equation}
and the system is quantized in the canonical quantization scheme by imposing
the following commutation relations
\begin{equation}
\lbrack a_{\vec{k}},a_{\vec{k^{\prime}}}]=0\hspace{0.5cm}[{a_{_{\vec{k}}%
}^{\dagger}},{a_{_{\vec{k^{\prime}}}}^{\dagger}}]=0\hspace{0.5cm}[{a}_{\vec
{k}},{a_{\vec{k}^{\prime}}^{\dagger}}]=\delta_{\vec{k}\vec{k^{\prime}}}.
\label{l}%
\end{equation}
With these commutation relations, $a_{\vec{k}}$'s and $a_{\vec{k}}^{\dagger}
$'s are annihilation and creation operators and the vacuum of the Lorentz
violation model in Minkowski space-time is defined by
\begin{equation}
a_{\vec{k}}\mid0_{_{LV}}^{^{M}}\rangle=0,\hspace{0.5cm}\omega\in%
\mathbb{R}
. \label{m0}%
\end{equation}

The above vacuum was defined with respect to a coordinate system in which the
components of $u^{\mu}$ were constant. We shall work in a coordinate system
which corresponds to the rest frame of a preferred inertial observer (aether).
In this case $u_{\mu}$ takes the form $(1,0,0,0).$ With respect to the aether
the dispersion relation (\ref{g}) and the normalization constant in (\ref{j})
are transformed to Eq.(\ref{a}) and $\omega^{-1/2}$ , respectively.

It is important to note that in what sense is the Lorentz invariance violated.
The Lagrangian (\ref{b})\ does not exhibit the invariance under particle boost
transformations \cite{reza1,reza2}. These are transformations of a physical
system (particles or localized fields) within a fixed coordinate system. It is
in this sense that Lorentz invariance is violated. Therefore the above vacuum
may not be considered as an invariant state under a particle Lorentz
transformation transforming a physical system within the rest frame of the
preferred observer (aether).


\section{Cosmological Model and Particle Creation}

\label{1011}

To show how particle creation occurs in the cosmological context, we use a
two-dimensional Robertson-Walker space-time with the line element
\begin{equation}
ds^{2}=dt^{2}-a^{2}(t)dx^{2}, \label{r}%
\end{equation}
and consider the generalization of the Lagrangian (\ref{b}) to this
cosmological space-time, namely%
\begin{align}
\mathcal{L}(\varphi,u_{\mu},\lambda)  &  =\frac{1}{2}\sqrt{-g}(g^{\mu\nu
}\nabla_{\mu}\varphi\nabla_{\nu}\varphi-\xi R\varphi^{2}+\alpha^{2}%
(D^{2}\varphi)^{2}\nonumber\\
&  +\lambda(1-u^{\mu}u_{\mu})), \label{m}%
\end{align}
where $D^{2}\varphi$ is the covariant spatial Laplacian \cite{7}(the covariant
analogue of (\ref{b2}))
\begin{equation}
D^{2}\varphi=-D^{\mu}D_{\mu}\varphi=-q^{\mu\nu}\nabla_{\nu}(q_{\mu}^{\tau
}\nabla_{\tau}\varphi),
\end{equation}
and $q_{\mu\nu}=-g_{\mu\nu}+u_{\mu}u_{\nu}$ with $g_{\mu\nu}$ corresponding to
(\ref{r}). $\xi$ is a coupling constant between the scalar field and scalar
curvature. The vector field $u_{\mu}$ is as a non-dynamical vector field to be
specified by the conditions of the theory. By introducing the conformal time
$\eta$, defined by $d\eta=dt/a(t)$, the metric (\ref{r}) takes the form
\begin{equation}
ds^{2}=c(\eta)(d\eta^{2}-dx^{2})\hspace{0.5cm},\hspace{0.5cm}c(\eta)=a^{2}(t).
\label{s}%
\end{equation}

The Lagrange multiplier $\lambda$ in (\ref{m}) imposes the following
constraint on $u^{\mu}$%
\begin{equation}
g_{\mu\nu}u^{\mu}u^{\nu}=1. \label{6}%
\end{equation}
In the context of the homogeneous and isotopic cosmological metric (\ref{s})
the vector field $u_{\mu}$ is taken as to satisfy the isotropic property of
cosmological metric and equation (\ref{6}). This leads to%
\begin{equation}
u_{\mu}\equiv(\sqrt{c(\eta)},0). \label{o2}%
\end{equation}
Thus $q_{\mu\nu}$ is%
\begin{equation}
q_{00}=q_{01}=q_{10}=0\hspace{0.5cm}and\hspace{0.5cm}q_{11}=c(\eta).
\label{o3}%
\end{equation}
Now setting the variation of the action $S=\int\,\mathcal{L}(\varphi)d^{4}x$
with respect to $\varphi$ equals to zero yields the equations of motion for
$\varphi$\ in the metric (\ref{s}) as follows
\begin{equation}
\square\varphi+\xi R\varphi-\frac{\alpha^{2}}{c^{2}\left(  \eta\right)  }%
\frac{\partial^{4}\varphi}{\partial x^{4}}=0. \label{4}%
\end{equation}
\begin{figure}[ptbh]
\centerline{\includegraphics[width=.45\textwidth,angle=0]{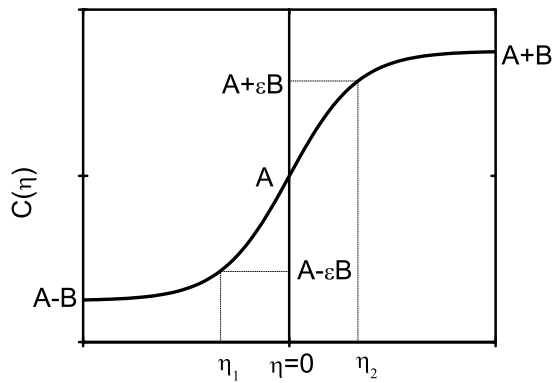}}
\caption{The figure shows $c\left(  \eta\right)  $ with respect to $\eta$. It
shows $c\left(  \eta=\pm\infty\right)  =A\pm B$ so $a\left(  \eta=\pm
\infty\right)  =\sqrt{A\pm B}$. Also it shows the time interval $\Delta
\eta=\eta_{2}-\eta_{1}$ that the proportion of total expansion that occurs in
this time interval is $\varepsilon$. We choose $\varepsilon=0.99.$}%
\label{110}%
\end{figure}

We want to quantize $\varphi$ in equation (\ref{4}) when $\xi=0$ (it is the
conformal coupling case in two dimension in absence of the Lorentz violation
term). For this purpose we will obtain the mode solutions $u_{k}(x)$ of
equation (\ref{4}).

In order to avoid the well known ambiguities in the particle concept in curved
space-time \cite{1}, we suppose that the space-time can be treated as
asymptotically Minkowskian in the remote past and future. We refer to the
remote past and future as the \textit{in} and \textit{out} regions ,
respectively. We take $c(\eta)$ as%
\begin{equation}
c(\eta)=A+B\tanh(\rho\eta), \label{u}%
\end{equation}
where $A$, $B$ and $\rho$ are some constants such that $2B$ and $\rho$
represent the amount and the rate of expansion in conformal time $\eta$,
respectively [see fig.(\ref{110})]. Then in the remote past and future the
space-time becomes Minkowskian since
\begin{equation}
c(\eta)\rightarrow A\pm B,\hspace{0.5cm}\eta\rightarrow\pm\infty. \label{v}%
\end{equation}
This kind of conformal scale factor first introduced by Bernard and Duncan in
\cite{BR-DU} and then has been used in many works
\cite{Setare,Tolley,Hamilton,Koks} for considering the cosmological particle creation.

We can solve Eq.(\ref{4}) and obtain $u_{k}(x)$ by the method of separation of
variables. The mode solution of this equation is%
\begin{equation}
u_{k}(\eta,x)=(2\pi)^{-\frac{1}{2}}e^{ik.x}\chi_{k}(\eta), \label{x1}%
\end{equation}
such that $\chi_{k}(\eta)$ satisfies the following equation
\begin{equation}
\frac{d^{2}}{d\eta^{2}}\chi_{k}(\eta)+({k}^{2}-\frac{\alpha^{2}}{c(\eta)}%
k^{4})\chi_{k}(\eta)=0. \label{8}%
\end{equation}
This equation can be solved in terms of hypergeometric functions. The
normalized modes which behave like the positive frequency Minkowski space
modes in the remote past $(\eta\longrightarrow-\infty)$ are
\begin{align}
{u_{k}}^{in}(\eta,x)  &  =(4\pi\omega_{in})^{-\frac{1}{2}}exp(ikx-i\omega
_{+}\eta\label{9}\\
&  -\frac{i\omega_{-}}{\rho}ln[(A+B)e^{\rho\eta}+(A-B)e^{-\rho\eta
}])\nonumber\\
&  \times F(1+i\omega_{-}/\rho,i\omega_{-}/\rho;1-i\omega_{in}/\rho
;z),\nonumber
\end{align}
and the modes which behave like positive frequency Minkowski modes in the
\textit{out} region as $\eta\longrightarrow+\infty$ are found to be
\begin{align}
{u_{k}^{out}}(\eta,x)  &  =(4\pi\omega_{in})^{-\frac{1}{2}}exp(ikx-i\omega
_{+}\eta\label{10}\\
&  -\frac{i\omega_{-}}{\rho}ln[(A+B)e^{\rho\eta}+(A-B)e^{-\rho\eta
}])\nonumber\\
&  \times F(1+i\omega_{-}/\rho,i\omega_{-}/\rho;1-i\omega_{out}/\rho
;1-z),\nonumber
\end{align}
where
\begin{equation}
z=\frac{1}{2}(A+B)\frac{\tanh(\rho\eta)+1}{A+B\tanh(\rho\eta)}, \label{y}%
\end{equation}
\bigskip and%
\begin{equation}%
\begin{array}
[c]{c}%
\medskip\omega_{in}=k(1-\frac{\alpha^{2}}{A-B}k^{2})^{\frac{1}{2}}\\
\medskip\ \omega_{out}=k(1-\frac{\alpha^{2}}{A+B}k^{2})^{\frac{1}{2}}\\
\omega_{\pm}=\frac{1}{2}(\omega_{out}\pm\omega_{in}).
\end{array}
\label{a1}%
\end{equation}

The energies $\omega_{in}$ and $\omega_{out}$ can be imaginary. But we limit
ourselves to the modes with real frequencies (the ordinary modes). The
ordinary modes ${u_{k}}^{in}(\eta,x)$ $\Bigl(u_{k}^{out}(\eta,x)\Bigr)$ are
orthonormal in the following conserved inner product%
\begin{align}
(  &  \varphi_{1},\varphi_{2})=\label{p}\\
&  -i\int_{\Sigma}\sqrt{-g}\varphi_{1}(x)(\overleftrightarrow{\partial}^{\mu
}-\alpha^{2}q^{\mu\nu}q^{\gamma\delta}\overleftrightarrow{{\partial_{\nu
}\partial_{\gamma}\partial}}{_{\delta}})\varphi_{2}^{\ast}(x)d\Sigma_{\mu
},\nonumber
\end{align}
where it is the generalization of inner product (\ref{h}) to curved space-time
and $\overleftrightarrow{{\partial_{\nu}\partial_{\gamma}\partial}}{_{\delta}%
}$ is defined in (\ref{n})$.$ Thus we may expand one ordinary solution of
Eq.(\ref{4}) with respect to ${u_{k}}^{in}(\eta,x)$ $\Bigl({u_{k}^{out}}%
(\eta,x)\Bigr)$ as%
\[%
\begin{array}
[c]{c}%
\bigskip\varphi(t,x)=\sum_{k}a_{k}^{in}u_{k}^{in}(t,x)+a_{k}^{\dagger in}%
u_{k}^{\ast in}(t,x)\ \ \ \ \ \ \ \ \ \ \omega_{in}\in\mathbb{R}\\
\varphi(t,x)=\sum_{k}a_{k}^{out}u_{k}^{out}(t,x)+a_{k}^{\dagger out}%
u_{k}^{\ast out}(t,x)\ \ \ \ \omega_{out}\in%
\mathbb{R}%
\end{array}
\]
and the second quantization is implemented in the same way as the Minkowski
space-time by the following commutation relations%
\begin{equation}
\lbrack{a}_{k}^{in},{a_{k^{\prime}}^{\dagger in}}]=\delta_{kk^{\prime}}%
\hspace{0.8cm}[{a}_{k}^{out},{a_{k^{\prime}}^{\dagger}}^{out}]=\delta
_{kk^{\prime}}. \label{q2}%
\end{equation}

As we mentioned before we have chosen a conformally flat expanding space-time
which is in the \textit{in} and \textit{out} regions Minkowskian. With using
the Eqs.(\ref{o2}) and (\ref{u}) we define the aether in the \textit{in} and
\textit{out} regions as a frame in which $u_{\mu}$ is $(\sqrt{A-B},0)$ and
$(\sqrt{A+B},0),$ respectively. The vacuums with respect to the aether in the
\textit{in }and \textit{out }regions are defined by%

\begin{equation}%
\begin{array}
[c]{c}%
\bigskip a_{\vec{k}}^{in}\mid0_{_{LV}}^{^{M}}\rangle^{in}=0\hspace
{0.8cm}\omega_{in}\in%
\mathbb{R}%
\\
a_{\vec{k}}^{out}\mid0_{_{LV}}^{^{M}}\rangle^{out}=0\hspace{0.8cm}\omega
_{out}\in%
\mathbb{R}
.
\end{array}
\label{q3}%
\end{equation}
\begin{figure}[ptbh]
\centerline{\includegraphics[width=.4\textwidth,angle=0]{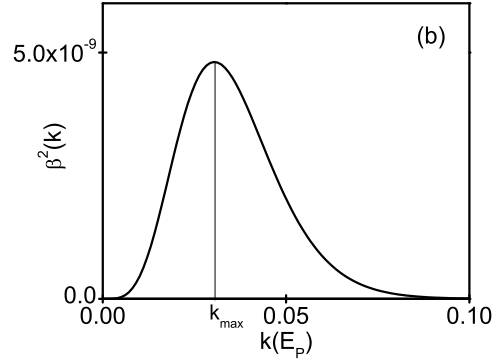}}
\vspace{0.3cm} \caption{The figure shows the spectrum of created \ particles
where $\sqrt{2B+1}=10^{3}$ and $\rho_{f}=10^{-5}$ so $\lambda\rho_{f}$
$=0.032$ $\ll1$ . (Note $\alpha=1^{-}$ , $\varepsilon=0.99$ and$~A-B=1$)}%
\label{120}%
\end{figure}

\begin{figure}[ptbh]
\centerline{\includegraphics[width=.40\textwidth,angle=0]{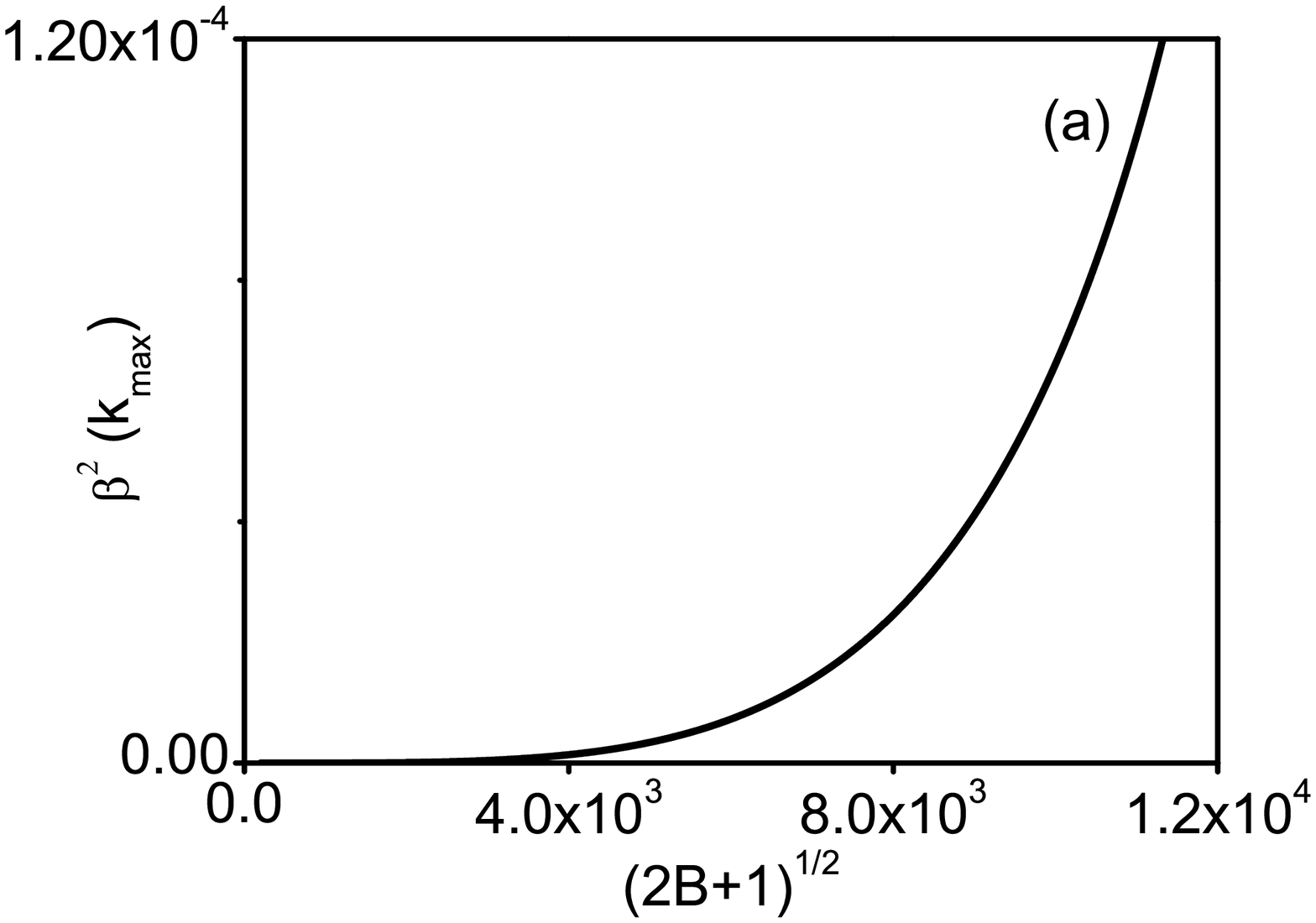}}
\vspace{-.5cm}
\centerline{\includegraphics[width=.40\textwidth,angle=0]{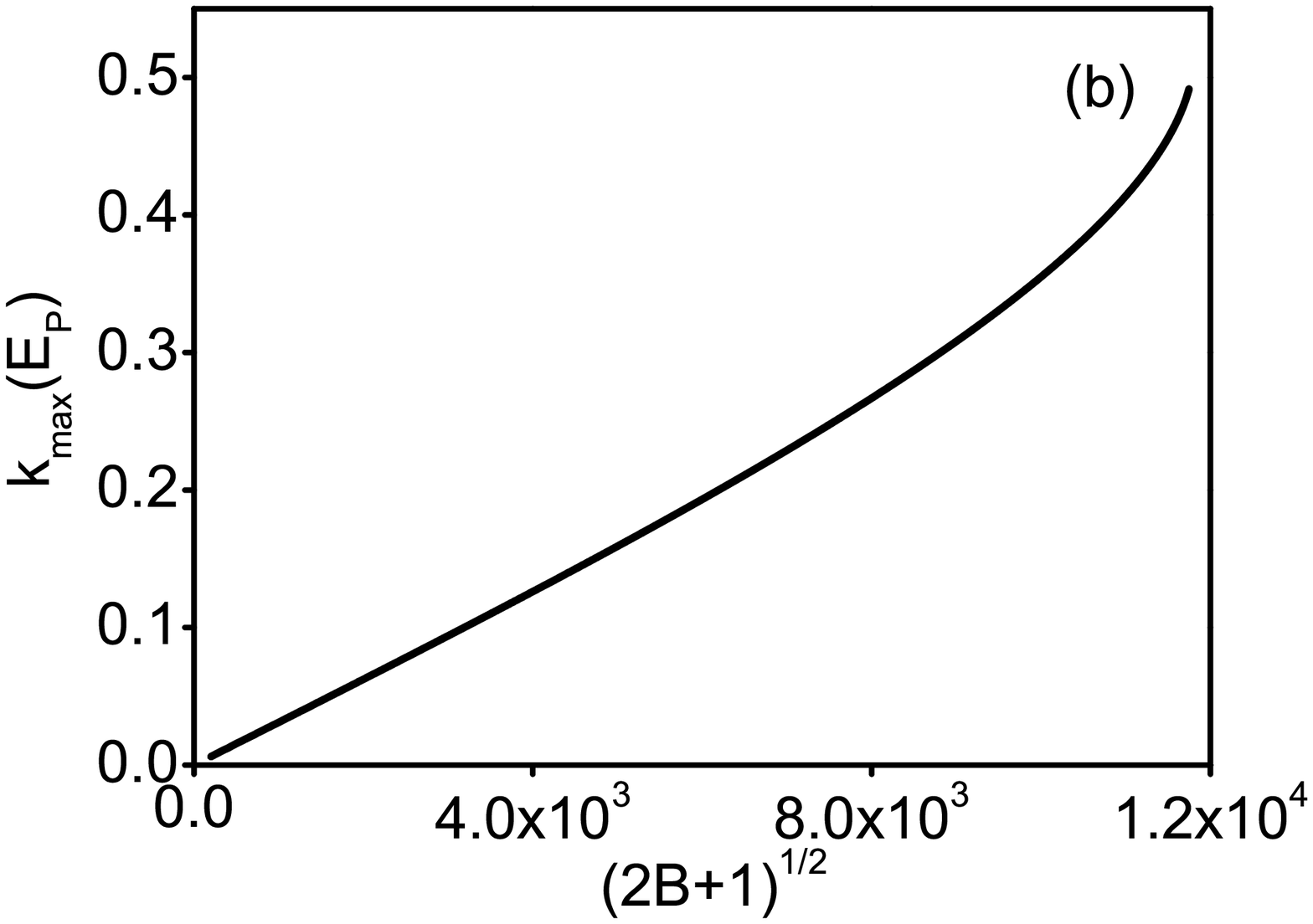}}
\vspace{-.5cm}
\centerline{\includegraphics[width=.40\textwidth,angle=0]{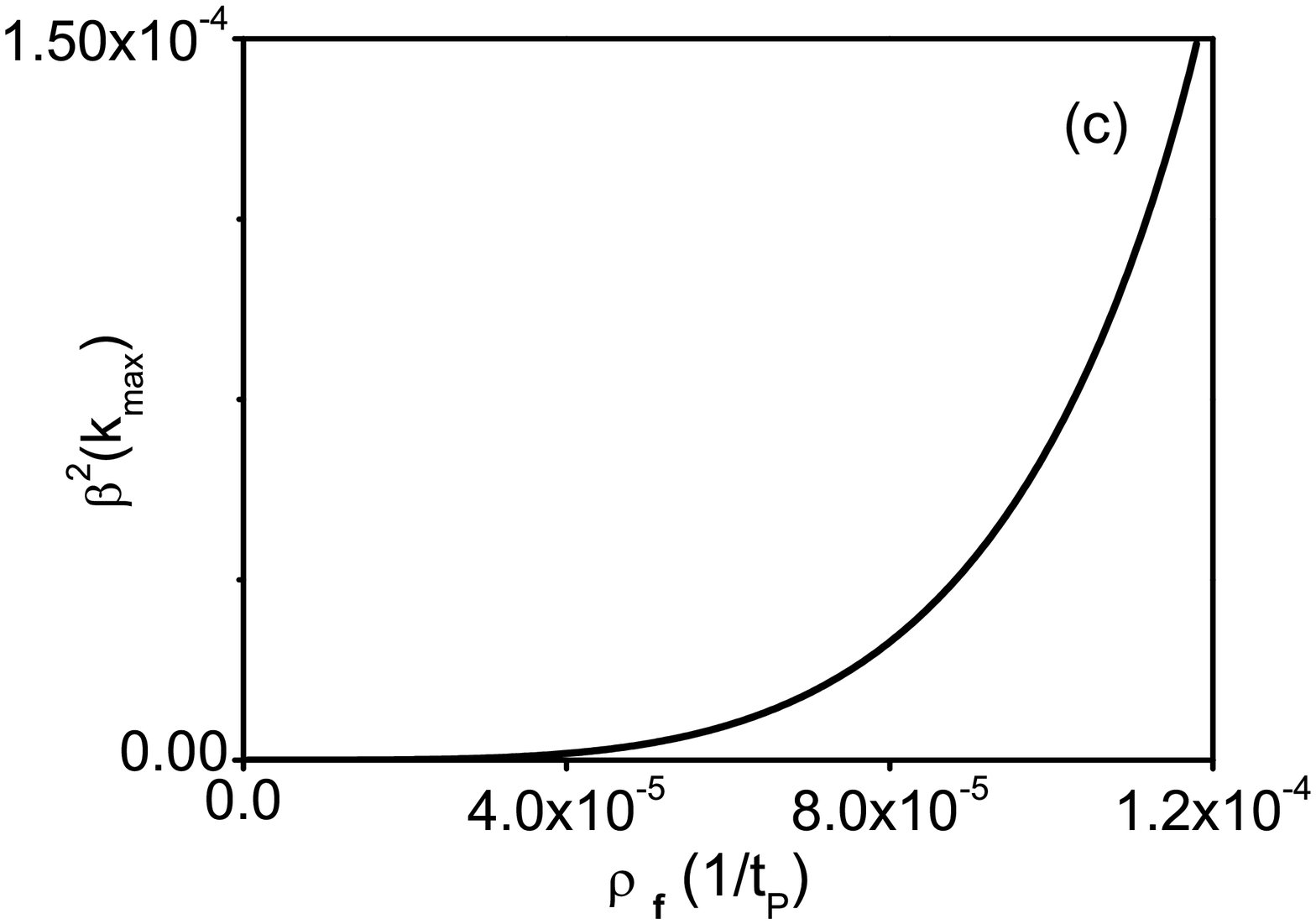}}
\vspace{-.5cm}
\centerline{\includegraphics[width=.40\textwidth,angle=0]{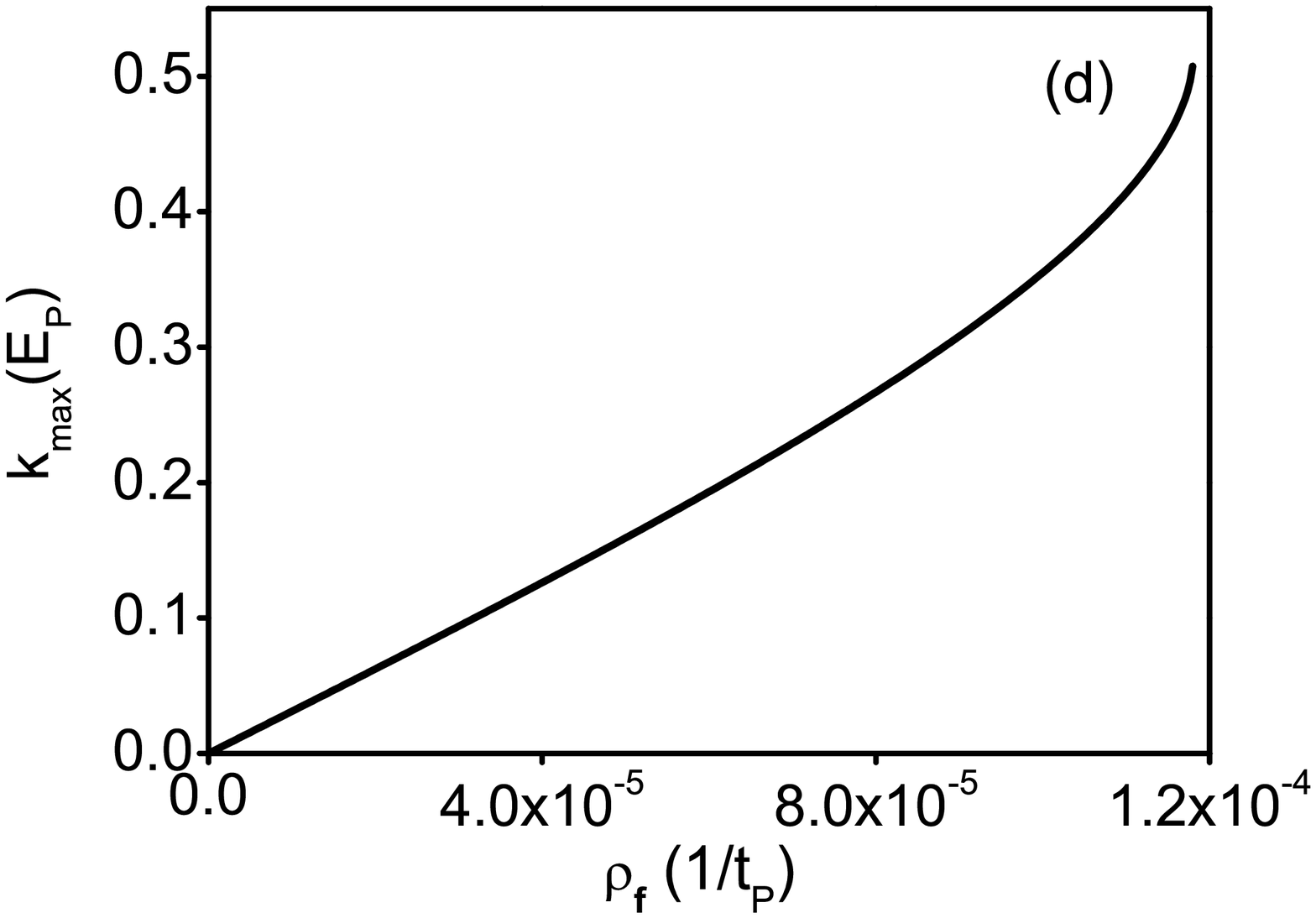}}\caption{The
above figures show the properties of the spectrum of created particles. (a)
and (b) show $\beta^{2}\left(  k_{\max}\right)  $ and $k_{max}$ versus the
final amount of the scale factor , $\sqrt{2B+1}$, \ where $\rho_{f}=10^{-5}$.
(c) and (d) show $\beta^{2}\left(  k_{\max}\right)  $ and $k_{max}$ versus the
rate of expansion in free fall frame, $\rho_{f}$, \ where $\sqrt{2B+1}=10^{3}%
$. (Note $\alpha=1^{-}$, $\varepsilon=0.99$ and$~A-B=1$)}%
\label{100}%
\end{figure}

Suppose the quantum field in the remote past resides in the state
$\mid0_{_{LV}}^{^{M}}\rangle^{^{in}}$and suppose we are working in the
Heisenberg picture. Thus in the \textit{out} region, the quantum field is also
in the state $\mid0_{_{LV}}^{^{M}}\rangle^{^{in}}$. But this state is not
regarded by the aether in the \textit{out} region as the physical vacuum, this
role being reserved for the state $\mid0_{_{LV}}^{^{M}}\rangle^{^{out}}$.

We want to calculate the number of particles detected in the \textit{out}
region in the state $\mid0_{_{LV}}^{^{M}}\rangle^{^{in}}$. If we use the
linear transformation properties of hypergeometric functions, we expand
${u_{k}^{in}}$ with respect to ${u_{k}^{out}}$ and obtain the following
relation
\begin{equation}
{u_{k}}^{in}(\eta,x)=\alpha_{k}{u_{k}^{out}}(\eta,x)+\beta_{k}{u_{-k}^{out}%
}(\eta,x), \label{b1}%
\end{equation}
where%
\[
\alpha_{k}=(\frac{\omega_{out}}{\omega_{in}})^{\frac{1}{2}}\frac
{\Gamma(1-(i\omega_{in}/\rho))\Gamma(-i\omega_{out}/\rho)}{\Gamma(-i\omega
_{+}/\rho)\Gamma(1-i\omega_{+}/\rho)}%
\]%
\begin{equation}
\beta_{k}=(\frac{\omega_{out}}{\omega_{in}})^{\frac{1}{2}}\frac{\Gamma
(1-(i\omega_{in}/\rho))\Gamma(i\omega_{out}/\rho)}{\Gamma(i\omega_{-}%
/\rho)\Gamma(1+i\omega_{-}/\rho)}. \label{c1}%
\end{equation}
as is well known, $\alpha_{k}$ and $\beta_{k}$ are the Bogolubov coefficients
and $\mid\beta_{k}\mid^{2}$ is equal to the number of particles\textit{\ }per
mode $k$ that is detected in the \textit{out }region (created particles),
\begin{equation}
\left\vert {\beta_{k}}\right\vert ^{2}=\frac{\sinh^{2}(\pi\omega_{-}/\rho
)}{\sinh(\pi\omega_{in}/\rho)\sinh(\pi\omega_{out}/\rho)}. \label{d1}%
\end{equation}

In the above relation $\left\vert {\beta_{k}}\right\vert ^{2}$ is dependent on
the parameters $\alpha$, $A$, $B$ and $\rho$. We set the first parameter
$\alpha$ in natural units $(\hbar=c=1)$ as $\alpha=1^{-}$ \footnote{The minus
in $1^{-}$ means approaching 1 from the left.} because we assume that LV
occurs in the scale of Planck energy. The three last parameters are related to
the conformal scalar factor $c(\eta)$ by (\ref{u}). These parameters determine
the spectrum of created particles with respect to the coordinates $\left(
\eta,x\right)  $ and metric (\ref{s}) through $\left\vert {\beta_{k}%
}\right\vert ^{2}$ in (\ref{d1}).

It is instructive to relate $\left\vert {\beta_{k}}\right\vert ^{2}$ to the
physical parameters in the free-fall frame which is determined by the
coordinates $\left(  t,x\right)  $ and the metric (\ref{r}). To do this we
first find the analogue of the parameters $A$, $B$ and $\rho$ in the free-fall
frame. The parameters, $A$ and $B$ have a good meaning in the free-fall frame
because as we have shown in fig.(\ref{110}), $\sqrt{A-B}$ and $\sqrt{A+B}$ are
initial and final scale factors in the free fall frame, respectively. For some
advantages we put $A-B=1$, it implies that the conformal time is the same as
the proper time in the remote past also the initial scale factor is $1$ and
the final scale factor is $\sqrt{2B+1}$. Now let us consider the analogue\ of
$\rho$. This parameter is proportional to the inverse of time interval of
expansion with respect to the conformal time. We show this by the following
argument. We define an effective time interval of expansion, $\Delta\eta.$ As
we have shown in fig.(\ref{110}), $\Delta\eta\left(  =\eta_{2}-\eta
_{1}\right)  $ is taken as a time interval such that the proportion
$\varepsilon$ of the total expansion is done in this interval. We set
$\varepsilon=0.99$ that means $99\%$ \ of total expansion occurs in
$\Delta\eta$. By this definition one can find ,from (\ref{u}), $\eta_{2}%
=-\eta_{1}=\tanh^{-1}\left(  \varepsilon\right)  /\rho$ and so%
\begin{equation}
\rho=\frac{2\tanh^{-1}\left(  \varepsilon\right)  }{\eta_{2}-\eta_{1}%
},\label{d1.5}%
\end{equation}
where $\eta_{1}$ and $\eta_{2}$ are the initial and the final time of the
effective expansion respectively, and $\tanh^{-1}\left(  0.99\right)  =2.64$.
The Eq.(\ref{d1.5}) shows $\rho\propto1/\Delta\eta$ and the proportional
constant is of order one. This relation makes it clear that $\rho$ is
proportional to the inverse of the time interval of expansion so it can be
interpreted as the rate of expansion. In analogy to $\rho$ we can define
$\rho_{f}$ in free falling frame such that it is the inverse of expansion time
interval in free falling frame, namely
\begin{equation}
\rho_{f}=\frac{1}{t\left(  \eta_{2}\right)  -t\left(  \eta_{1}\right)
}.\label{d3}%
\end{equation}
For evaluating $\rho_{f}$ in the above relation we find the relationship
between $t$ and $\eta$ from (\ref{u}) and $dt=\sqrt{c\left(  \eta\right)
}d\eta$. We get%
\begin{align}
t\left(  \eta\right)   &  =\frac{\sqrt{2B+1}}{\rho}\tanh^{-1}\left(
\frac{\sqrt{B\left[  1+\tanh\left(  \rho\eta\right)  \right]  +1}}{\sqrt
{2B+1}}\right)  \nonumber\\
&  +\frac{1}{2\rho}\ln\left(  \frac{\sqrt{B\left[  1+\tanh\left(  \rho
\eta\right)  \right]  +1}-1}{\sqrt{B\left[  1+\tanh\left(  \rho\eta\right)
\right]  +1}+1}\right)  +C,\label{h1}%
\end{align}
where $C$ is a constant that sets the initial values of $\eta$ and $t$. From
(\ref{h1}) together with (\ref{d3}) we find $\rho_{f}=\rho/\lambda$ where
$\lambda$ is a function of $B$ and $\varepsilon$, namely
\begin{align}
\lambda &  =\sqrt{2B+1}\tanh^{-1}\left(  \frac{\sqrt{B\left(  1+\varepsilon
\right)  +1}}{\sqrt{2B+1}}\right)  \nonumber\\
&  -\sqrt{2B+1}\tanh^{-1}\left(  \frac{\sqrt{B\left(  1-\varepsilon\right)
+1}}{\sqrt{2B+1}}\right)  \nonumber\\
&  +\frac{1}{2}\ln\left(  \frac{\sqrt{B\left(  1+\varepsilon\right)  +1}%
-1}{\sqrt{B\left(  1+\varepsilon\right)  +1}+1}\right)  \nonumber\\
&  -\frac{1}{2}\ln\left(  \frac{\sqrt{B\left(  1-\varepsilon\right)  +1}%
-1}{\sqrt{B\left(  1-\varepsilon\right)  +1}+1}\right)  .\label{h2}%
\end{align}
From (\ref{h2}) we find that for sufficiently large amount of $B$, $\lambda$
can be approximated by $\left(  2B\right)  ^{\frac{1}{2}}$. Substituting
$\rho$ by $\lambda\rho_{f}$ in (\ref{d1}) we get $\left\vert {\beta_{k}%
}\right\vert ^{2}$ in term of the physical parameter $\rho_{f}$ in the free
falling frame. We can approximate (\ref{d1}) in the limit $\lambda\rho_{f}%
\gg1$ by%
\begin{equation}
\left\vert {\beta_{k}}\right\vert ^{2}=\frac{(\omega_{out}-\omega_{in})^{2}%
}{4\omega_{in}\omega_{out}},\label{h3}%
\end{equation}
where in the region $B\gg1$ and $k\neq0$ the above relation takes the form%
\begin{equation}
\left\vert {\beta_{k}}\right\vert ^{2}=\frac{\left(  1-\sqrt{1-\alpha^{2}%
k^{2}}\right)  ^{2}}{4\sqrt{1-\alpha^{2}k^{2}}},\label{h4}%
\end{equation}
which implies that $\left\vert {\beta_{k}}\right\vert ^{2}$ is independent of
$\lambda\rho_{f}$.

In the opposite limit where $\lambda\rho_{f}$ is small the spectrum shows an
interesting behavior. The figure(\ref{120}) shows the spectrum of created
particles for $\sqrt{2B+1}=10^{3}$ and $\rho_{f}=10^{-5}$ so $\lambda\rho_{f}$
$=0.032$. It shows that particles are not created at low and high momenta.
Most particles are produced with momenta around$\ k_{max}$. In figure
(\ref{100}) we show the properties of this spectrum. Figures (3,a) and (3,c)
show the variations of the number of created particles at momentum $k_{max}$
with respect to the amount , $\sqrt{2B+1}$, and the rate , $\rho_{f}$,of the
space-time expansion respectively. With increasing the amount and the rate of
expansion the number of created particles increases. Figures (3,b) and (3,d)
shows $k_{max}\ $with respect to $\sqrt{2B+1}$ and $\rho_{f}$. We see that
with increasing the amount and the rate of the expansion, the particles with
higher momenta are created.

\begin{acknowledgments}
E. Khajeh and N. Khosravi thank H. Salehi for encouragements. E. Khajeh thanks
S. Jalalzadeh for useful discussions and research office of Shahid Beheshti
University for financial supports. The authors thank an unknown referee for
useful comments.
\end{acknowledgments}

\end{document}